# Web Based Monitoring in the CMS Experiment at CERN

**William Badgett** [a], **Laura Borrello**[b], **Irakli Chakaberia**[c], **Dominique Gigi**[d], **Youngkwon Jo**[e], **Juan Antonio Lopez-Perez**[a], **Kaori Maeshima**[a], **Sho Maruyama**[a], **James Patrick**[a*], **Valdas Rapsevicius**[a], **Aron Soha**[a], **Balys Sulmanas**[a], **Zongru Wan**[c]

[a] *Fermi National Accelerator Laboratory,*
*Batavia, IL, USA*
[b] *University of Wisconsin,*
*Madison, WI, USA*
[c] *Kansas State University,*
*Manhattan, KS, USA*
[d] *CERN,*
*Geneva, Switzerland*
[e] *Korea University,*
*Seoul, KR*

*E-mail*: patrick@fnal.gov

ABSTRACT: The Compact Muon Solenoid (CMS) is a large and complex general purpose experiment at the CERN Large Hadron Collider (LHC), built and maintained by many collaborators from around the world. Efficient operation of the detector requires widespread and timely access to a broad range of monitoring and status information. To this end the Web Based Monitoring (WBM) system was developed to present data to users located anywhere from many underlying heterogeneous sources, from real time messaging systems to relational databases. This system provides the power to combine and correlate data in both graphical and tabular formats of interest to the experimenters, including data such as beam conditions, luminosity, trigger rates, detector conditions, and many others, allowing for flexibility on the user side. This paper describes the WBM system architecture and describes how the system was used during the first major data taking run of the LHC.

---

[*] Corresponding author.

# Contents

## 1. Introduction

The Compact Muon Solenoid (CMS) [1] is a general purpose detector at the Large Hadron Collider (LHC) [2] at CERN built to study proton-proton, proton-heavy ion, and ion-ion collisions at center of mass energies up to 14 TeV. The operation of CMS involves tens of millions of sensor channels, substantial supporting infrastructure, and complex trigger and data acquisition (DAQ) systems. The vast and diverse information to monitor includes detector and environmental conditions, DAQ status, run configuration, trigger rates, luminosity, and accelerator parameters. Extensive monitoring is important for efficient and high quality data taking. Major subsystems such as the CMS tracker provide dedicated monitoring applications for shift personnel at the experiment site. These are often not accessible from off-site. Also each application usually focuses on a specific part of the online system. The CMS collaboration is truly global, with over 3000 collaborators from over 175 institutions from over 40 countries. With such distributed expertise and a variety of data sources, there is a need for unified remote access to monitoring data in order to better identify and correct issues with data taking. It is also valuable to give remote experts easy access to this data so they may help in diagnosing problems.

The CMS Web Based Monitoring (WBM) system is a set of hardware and software that aggregates data from various online monitoring sources and makes them available to a set of web based applications accessible to any authenticated user anywhere. The various applications give access to a summary of the current status of the experiment, as well as convenient access to historical data. A single application can easily include information from multiple online sources. These are used by shift crew members, detector subsystem experts, operations coordinators, and those performing physics analyses. The WBM tools are extensively used by the CMS Run Coordination and other upper level managers for generating reports of data taking achievements, records, and efficiency. The tools have become important in managing data taking operations for the CMS detector. The type of monitoring described in this paper is complementary to other data quality monitoring tools that are based on event data [3]. The goal of CMS WBM is to provide an interactive suite of tools to support effective central data taking that is easy to use, secure, flexible, and maintainable.

This paper will describe the WBM infrastructure, core applications providing overall summary information and also some more specialized tools, some of the detector subsystem monitoring applications, experience with the system during data taking, and future plans.

## 2. System Infrastructure

A simplified view of the WBM system architecture is displayed in Figure 1. The following sections provide details of the system components. A set of computers aggregates data from various CMS detector subsystems and the LHC accelerator. They write this information to an online Oracle database, provide some data directly to online real-time applications and insert some information into the event data stream. The data are accessed by a number of web based applications. The local web servers provide information to the crew at the experiment site. The applications are accessible outside of the experiment network via a proxy server on the CERN general network. To reduce the load on the online web servers, dedicated offline web servers operate on a copy of the online database and on data from the CMS offline database. These services are accessible for collaborators worldwide. The WBM system is compliant with CERN security standards.

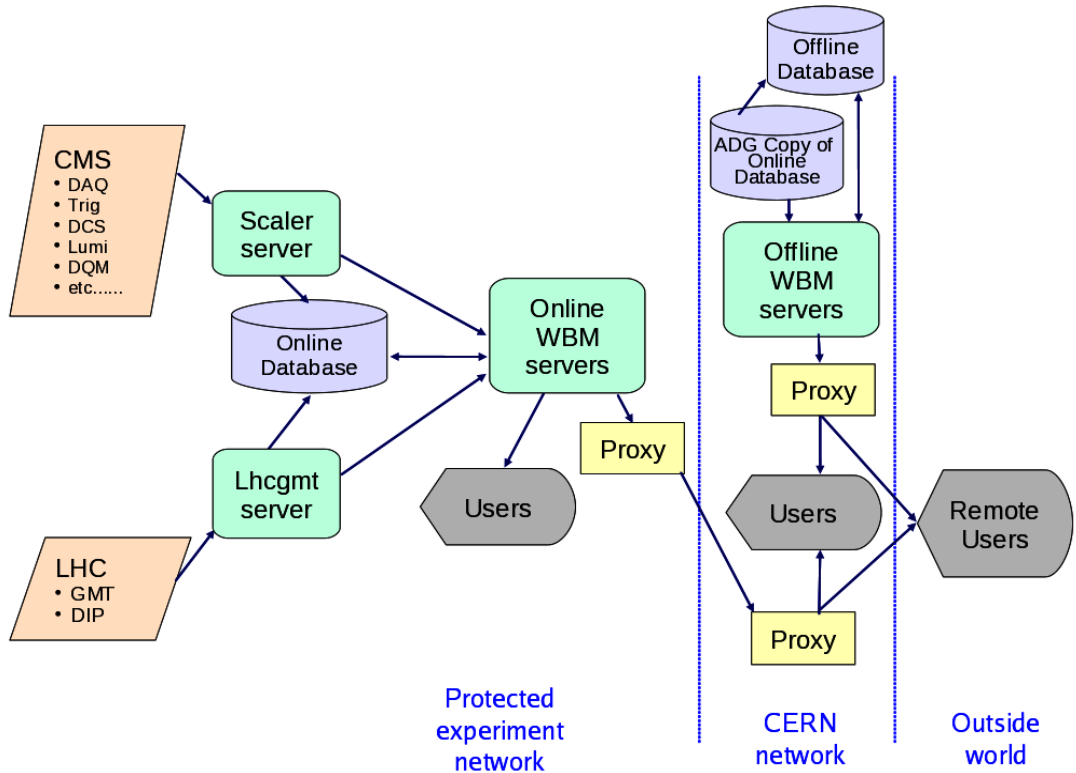

Figure 1. Overview of the dataflow in the CMS Web Based Monitoring system

## 2.1 Sources of Information

Information is obtained from the CMS detector subsystems and the LHC accelerator by a combination of specialized hardware and also via various software protocols over ethernet. Data are made available in real time and also logged by the WBM system in a database with time stamps for later retrieval. Data saved by the WBM system as well as directly by subsystems may be retrieved from the WBM server in XML format on the experiment protected network.

The LHC General Machine Timing (GMT) system [4] provides synchronous event timing and brief status information over dedicated high bandwidth serial links throughout the accelerator complex. The GMT signal from the LHC is made available to CMS through a dedicated PC within the WBM subsystem at the CMS experiment site. That PC has a hardware interface card that receives the incoming GMT events and fields indicating the current status of the LHC, including beam intensities, energy, safety and permit flags, and other information. The PC runs CMS specific software including an interface to the hardware card. The WBM software receives this information and converts it into other formats to provide monitoring inside and outside the CMS experiment protected network. It also recognizes important changes in the machine status such as the start and end of a period of stable colliding beams (called a Fill at the LHC) and writes this information to the database.

More extensive accelerator related information is also exchanged between the LHC equipment and the experiments using the Data Interchange Protocol (DIP) [5] [6]. This system allows relatively small amounts of real-time data to be exchanged between very loosely coupled heterogeneous systems. DIP is intended for systems that do not require very low latency. The data are assumed to be mostly summary data rather than low-level parameters from the individual systems. Some information relevant to the machine such as luminosity and beam losses measured by CMS are published by CMS. The data can be accessed via software by subscribing to the published trees of information. The servers publishing the information are located in different protected networks, such as the LHC technical network and the CMS experiment protected network, and are not trivially accessible across networks, especially from outside of CERN. Different parts of the WBM software subscribe to LHC and CMS DIP information branches, to monitor and display the data. Information from the Detector Control System (DCS) [7] [8] such as high voltage status and various environmental information is also published by that system via DIP. This information is also logged in the database by the WBM software where it can be viewed by non-real time applications.

Important real-time data, such as the status of the trigger and DAQ systems, are provided via the monitoring system of the online data acquisition software framework XDAQ. XDAQ [9] is a software platform designed at CMS specifically for the development of distributed data acquisition systems. WBM for example acceses the XDAQ monitoring information from the trigger system to monitor and display real-time trigger information for shift-takers and experts. Such information is also stored by the WBM code in the database for later debugging and validation of the data. By examining various rate-counter information, starts and stops in data taking are noted and written to the database. This information is used by other WBM services that account for operational down time.

Online luminosity information measured by CMS subdetectors is obtained from the luminosity subsystem and also logged to the database. This information is used to mark blocks of delivered luminosity, approximately 23 seconds in duration, called Luminosity Sections. The detector high voltage status is recorded in the database for each of these sections. Luminosity sections where some part of the detector is not functioning properly may be excluded from some offline physics analyses. These sections are long enough to allow a precise determination of the luminosity but short enough so that data loss is minimized when the detector is not fully functional.

### 2.2 Databases

The CMS online database contains a wealth of information regarding the current and past states of the CMS detector written by a variety of sources in addition to the information logged by the WBM system [10]. It is an Oracle database, with several servers working within the protected CMS experiment network. The stored data is copied to a database server on the general network via Oracle Advanced Data Guard [11]. A natural method to convey and display this information is a web server, and the WBM server was proposed and implemented to allow globally useful, but carefully controlled, access to the data for people at the CMS site, as well as for people at remote locations.

### 2.3 WBM Computers

The WBM infrastructure includes the following machines running the 64-bit version of Scientific Linux CERN (SLC) [12]. The main use for each machine is also listed.

- Lhcgmt server: continuously receives LHC GMT signals from the accelerator and DIP information from the LHC and CMS. It logs this data in the database, and also records periods of stable beams. This data is also republished via DIP to the rest of CMS.
- Scaler server: receives real-time scaler quantities, such as trigger rates, via the data acquisition monitoring system and logs them to the database. It also uses changes in the DAQ and BEAM status to automatically log information about periods when CMS is not recording data. It is equipped with interfaces to the Trigger, Timing and Control System (TTC) and the DAQ system which allows it to inject various data into the event data stream. In addition to scaler information the system also inserts its own unique data including timing information, the bunch crossing, and orbit number history. This event stream data is then used by the High Level Trigger (HLT) [13], the Data Quality Monitoring (DQM) and some CMS subdetectors including the Tracker

- SCAL Function Manager: interfaces the WBM subsystem to the central DAQ run control software and controls the functioning of the lhcgmt and scalers machines.
- WBM web servers: currently two production WBM servers are located within the CMS protected network, and another offline WBM server is on the CERN Network.
- WBM development and spare machines: several machines are used to develop and test new services and new releases before deployment to production in both the P5 and offline environments. These machines also function as working spares.

**2.4 Software Implementations**

On the server side, the WBM server runs an Apache daemon to respond to HTTP requests. Most of the user interactive work is handled through Apache/Tomcat [14] using a Java servlet architecture. The main Apache daemon serves a small number of static pages and forwards requests to the Tomcat engine. JDBC and SQL are used to read data from the database. Some servlets subscribe to real time DIP data on request. While much of the code is written in Java, some data analysis and production of some static plots is performed by the C++ based ROOT package [15] with output available as PNG images or ROOT files that can be saved for further analysis. JFreeChart [16] is used to produce some PNG plots where the other formats are not required.

On the client side, data are presented in web pages using basic HTML, CSS, and JavaScript that is supported by all browsers. AJAX and XML are used to do automatic refreshing of real-time displays. Some client services are in the form of Java applets. In the second half of the 2012 LHC run, some of these were migrated to HTML5, jQuery [17], and Highcharts [18] which avoid various security and other issues with Java applets. We have found Highcharts to be an excellent package. It supports a variety of chart types, supports interactive zooming and mouse-over display of data values, and has good performance. The plan is to eventually replace all applets in this manner.

**2.5 Documentation, Code Repository, and Support**

The WBM project is documented primarily using Twiki [19] pages, and also in some of the shift taker instructions and internal collaboration notes. A subset of the pages that are hosted outside the experiment protected network are regularly copied into the protected network. This is done so that people are able to access key documentation and continue running the experiment even in rare instances when the network connection is interrupted.

Subversion [20] repositories are used for WBM software and other web page content related to CMS operations. For the main WBM services, development code is kept in the trunk of the repository and tags are created for each new release.

Issue tracking and problem reporting are carried out using a Savannah system [21]. Developers use this to keep track of current tasks, while WBM users use it to provide feedback, report possible problems, and request new features. User support and internal communications are also performed through CERN mailing lists. An on-call person is located at CERN who users, shift crews, and run coordination may contact directly in case of urgent issues.

**2.6 Reliability Features**

As the WBM system is important for CMS operations, a number of steps are taken to ensure the system is reliable. The computing infrastructure is monitored as part of the general CMS online system. In addition, the WBM system employs cronjob scripts to check the tomcat log files for errors and perform a restart if needed. These also look at a few more WBM specific things and notify experts if problems are found. The databases are monitored by the CERN IT Department and periodic reports are issued. The offline WBM machines are also monitored via CERN IT custom monitoring systems.

## 3. WBM Services

The WBM services are accessible through a simple web page, shown in Figure 2 that provides a single point of entry. A listing of the services is given in Table 1, and they are described in more detail in the following sections. There are monitoring services that focus on the both real-time and historical status of the detector and accelerator. Other tools track data taking efficiency, and perform other functions helpful for detector operations.

**CMS Web Based Monitoring online**

**Subdetectors WBM**
ECALSummary
DTSummary
RPCSummary
HCALHome
CSCSummary
BRMSnapshots
BCM1F Bunch Info
TriggerModes
TrackerTools
PixelHome
$S^3$ ScreenSnapShots

**Core Services**
RunSummary [24h] [24h&1+trig]
RunTimeSummary [LHC Fills] Deadtime
FillReport [Latest Fill] DataSummary
LumiScalers | Automatic Fill eMails
DQM Run Registry | Online DataQualityLogger $\beta$
TriggerHistory | TriggerRunListing
TriggerRates [Pre-DT L1] [Post-DT L1] [HLT]
LastValue | ConditionBrowser [iPlot]
MagnetHistory | CurrentBunches | BunchFill
LhcMonitor | LHCStatusDisplay | BLM | BPM | DIP
LhcCollimators | AbortGaps
ShiftAccountingTool
PageZero | CMS Page 1

**Links**
Online DQM GUI
FNAL ROC
Commissioning & Run Coordination
CMS Twiki: OnlineWB TriDAS
CMS Online
Shift eLog
Snappy eLogViewer
LHC Page 1

WBM Twiki Page | WBM Savannah | WBM Support & Contact
Last modified 2014-04-08 16:44:41 UTC

Figure 2. CMS Web Based Monitoring main page

| Page Zero | Real-time information from the LHC, trigger, data acquisition, and luminosity on a single page |
|---|---|
| Page 1 | Public page showing high level status of the LHC and CMS detector with plots of delivered and recorded luminosity |
| FillReport | Information and plots for LHC fills with links to associated data taking runs |
| RunSummary | Detector and trigger configuration, and data collected for data taking runs with links to more detailed information. |
| DataSummary | Plots and tables of luminosity and efficiency available by day, week, and year; and performance records for various time periods |
| TriggerRates | Display and plot real-time trigger rate information and warn of problematic conditions |
| TriggerHistory | Plot trigger cross sections versus luminosity and fit to proper functional form |
| LHC Status | Variety of information about the status of the LHC from the LHC, including beam conditions and current and machine state |
| LHC Status Display | Graphical display of LHC state in the context of possible states |
| LHC Information | Status of machine from CMS detector measurements such as beam backgrounds and losses |
| Deadtime Summary | Monitor detector dead time and its origin |
| LastValue | Current status of voltages and temperature of all CMS subdetectors |
| ConditionBrowser | Plot any values versus time, or any two time series versus each other for many detector and environment and beam conditions |
| Subdetector WBM | Tools to view the status and configuration of CMS subdetectors |
| Run Time Logger | Record experiment operational efficiency and reasons for down time and provide reports |
| Run Registry | Track quality of event data for monitoring and data certification |
| WBM Alarm Notification | Centralized alarm notification |
| Shift Accounting Tool | Provide statistics for official CMS shifts by person and institution |
| Snappy eLog Viewer | Lightweight and flexible tool to view the electronic logbook |
| DIP Logger | Receive and log LHC timing events and detector scaler information. Inject some of this data into the event stream |

Table 1. Summary of WBM services

## 3.1 Monitoring Services

The PageZero and CMS Page1 services provide a basic one screen non-interactive overview of the detector and accelerator status, and contain links to other WBM services that provide more details. PageZero summarizes the status of CMS and the LHC in various tables, while CMS Page 1 (see Figure 3) gives a high level summary which includes beam conditions, a plot of delivered and recorded luminosity, detector high voltage and readout status, trigger rate, and comments from both the LHC operators and the CMS shift leader. CMS Page1 is a public page and is accessible to anyone.

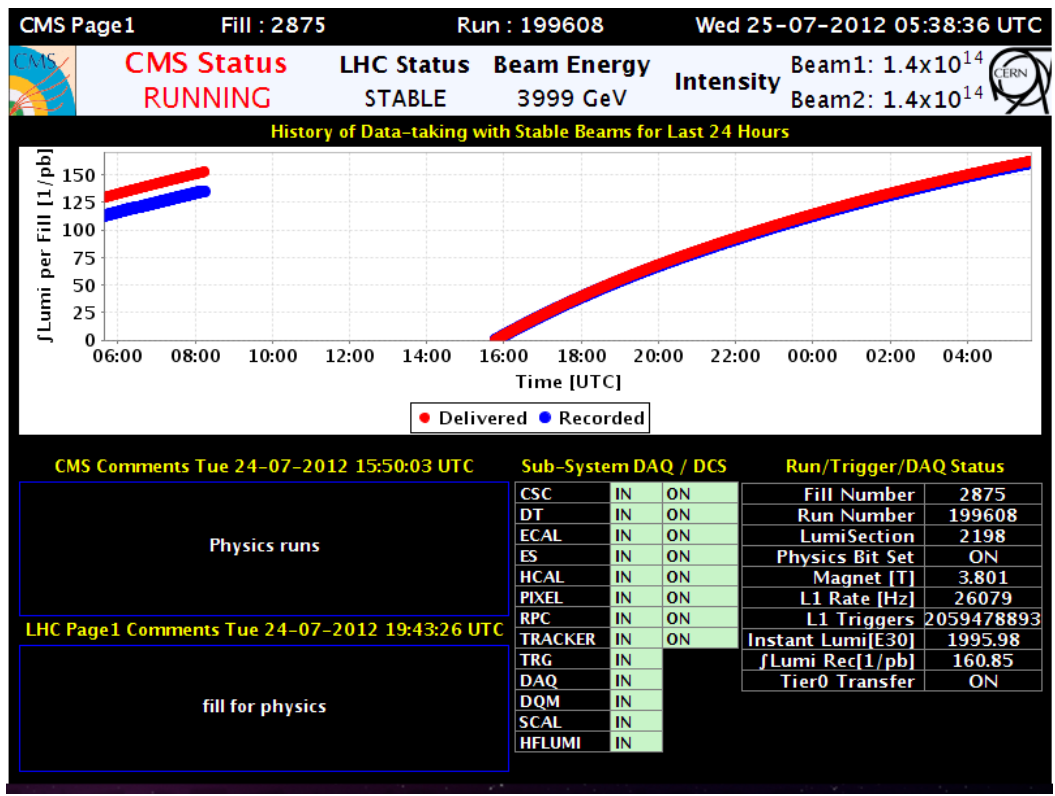

Figure 3. CMS Page 1 display. The difference in delivered and recorded luminosity in the left fill primarily represents down time earlier in the fill. For the fill on the right the smaller difference is due almost entirely due to dead time during data taking.

The LhcStatusDisplay service indicates the current mode of the LHC such as "Stable Beams", "Ramp", or "Beam Dump", in the context of the possible states. The LhcMonitor service presents more detailed information about the LHC parameters in tabular form. It includes the most recent status telegram from the GMT system, a list of the most recent GMT events, and basic parameters for each beam. The BunchFill service provides a visualization of the complex bunch patterns that populate the LHC accelerator rings. The chart and tables are available for the current collision period as a dynamic, real-time display, or for previous running periods. Additional pages display the real-time status of the LHC collimators, abort gap status, beam position monitors and beam loss monitors. This information is provided via DIP and the GMT link from the accelerator.

The TriggerRates service (Figure 4) displays trigger counts, rates, and configurations, either in real time for the current run, or historically from the database. Both Level 1 rates, as well as High Level Trigger (HLT) rates may be plotted based on either real time or detector live time. It includes color-coded indicators for triggers that operate outside their preset ranges, and includes check boxes to select which trigger rates are plotted. The valid ranges are computed using the TriggerHistory service described in section 3.2. The plots allow interactive zooming and display values when the mouse is over a data point. The Deadtime Summary shows the data taking dead time for either some time period or set of fills. It shows the contributions from limitations on data link speed and buffer capacity creating back-pressure in the data acquisition system, calibration events taken while running, rules that limit the frequency of triggers, and intentional pauses in data taking.

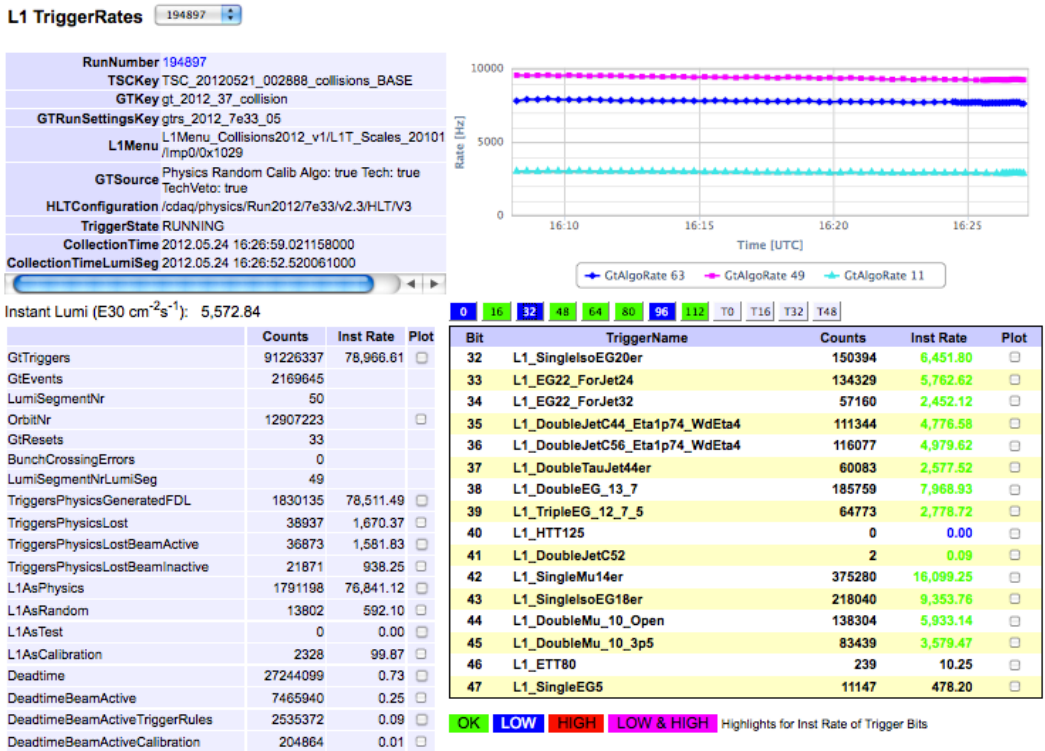

Figure 4: Trigger Rates page. This shows global configuration and rate information, and detailed information about specific triggers. Here the monitor has flagged a trigger with a low rate. This one has a large prescale factor applied before it fires.

The FillReport displays information about the current LHC fill or past fills. The report contains tables and plots of luminosity, backgrounds, accelerator information, trigger rates etc. Included is a list of data taking runs associated with the fill with links to more information about them. Examples are shown in Figures 5 and 6. This service is able to send e-mails to subscribed users with notifications of the start and end of the fill as well as information on the filling scheme and integrated luminosity acquired. The FillReport is heavily used by the Run Coordinators and shift crew to monitor and follow up current and recent issues with the experiment operation.

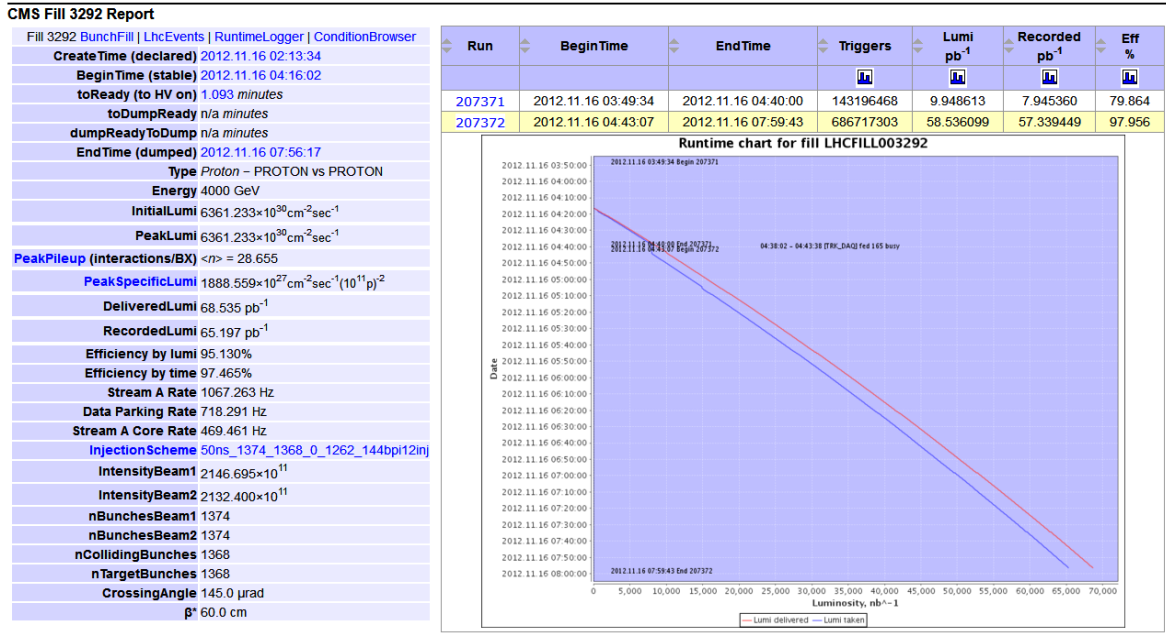

Figure 5. FillReport summary for a fill showing configuration and statistics for the fill, a table of runs taken and a plot of delivered and recorded luminosity versus time.

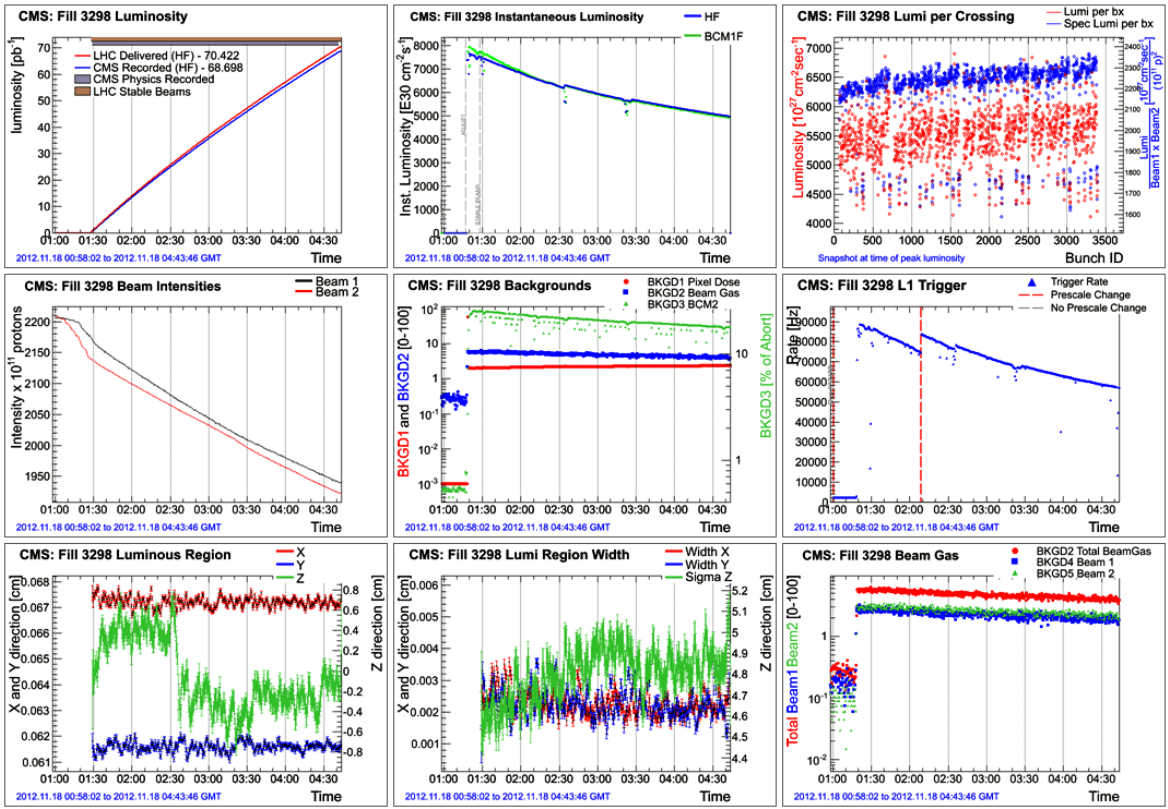

Figure 6. Selection of FillReport summary plots

The RunSummary service displays information about CMS data taking runs. It includes information about the detector and trigger configuration, and data collected. Runs are periods of constant detector configuration taken during LHC fills. The RunSummary is searchable and provides links to other WBM services providing more detailed information about the time evolution during the run and to data quality plots based on event data. Examples of useful links are pages with complete Level 1 and HLT trigger information, including details of the trigger configuration and on-demand plots of rates. Another example is the monitoring of system dead time, detailing which sources within the detector are causing dead time. Dead time occurs when the data buffers in some subsystem are full such that new triggers must be inhibited. Links to LHC fill information and statistics are available. Sub-detector specific pages are also conveniently linked from this run summary page. RunSummary thus forms a top level portal into the detailed WBM information for data taking runs. Figure 7 shows an example of the main Run Summary page and several subdisplays.

Figure 7. RunSummary usage example illustrating links to more detailed information

The DataSummary service displays brief summary information and plots of instantaneous and integrated luminosity and data taking efficiency. These can be generated by day, week, or year. By default these are shown for the most recent time period but may be generated for earlier time periods as well. The page contains links to the more detailed fill and run reports, and also to tables of records for luminosity and efficiency for the run.

The TriggerHistory service plots trigger cross sections as a function of luminosity. This dependence can be non-linear due to an increasing number of interactions per crossing as the luminosity rises. A service allows fits to be performed and the results used to predict and monitor rates for future runs, and also to design future trigger configurations. Static plots are generated daily. The fit results are used to generate allowed ranges for each trigger. These are used to provide notification during data taking when the observed rate is significantly different than the fit prediction. Figure 8 shows a fit of the trigger cross section versus instantaneous luminosity for a sample trigger created using this tool.

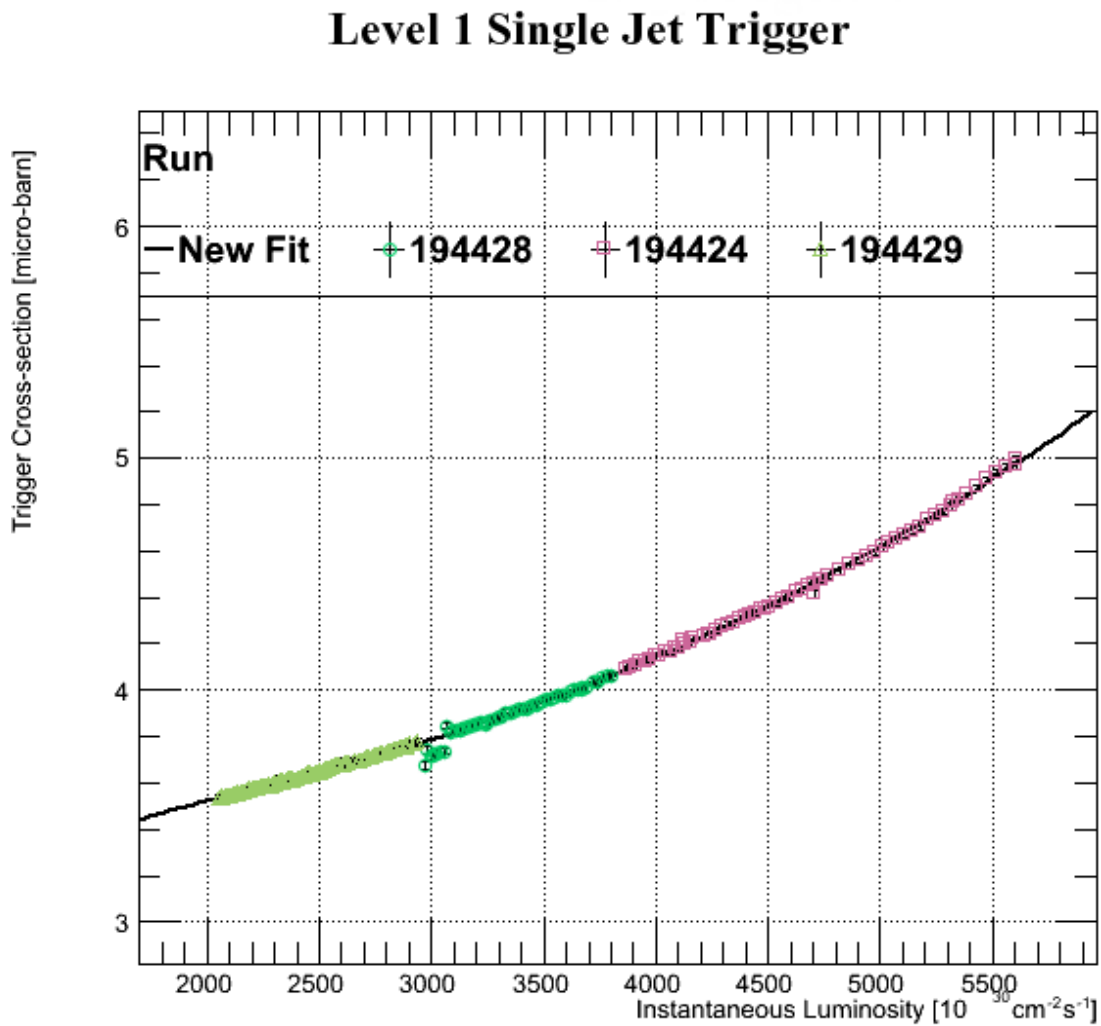

Figure 8. Trigger History Plot with fit to the cross section versus luminosity

The LastValue service consists of interactive displays for sub-detectors. It includes clickable schematic representations of the detector geometries and a browsable tree. Either interface can be used to drill down to obtain details of the detector operating parameters (voltages, currents, etc.) and environmental conditions (temperatures, humidity levels, etc.). Quantities can be plotted as a function of time, and the resulting plots and underlying data are downloadable.

The ConditionBrowser is an application for visualizing any of the more than 4000 quantities that are stored in the database in the form of values and time stamps. Trends over time can be plotted for any given time range, where the time stamps may be actual times, or a range of run numbers or luminosity sections. The implementation uses metadata, which describe where to find the values and time stamps for each quantity, as well as other optional default parameters to use when plotting. The underlying quantities are stored by different subsystems in diverse database schema, so the ConditionBrowser provides a clean user interface, and a layer of abstraction between the user and complex data queries, to give simple and uniform access to all of the variables. An example of two variables plotted as a function of time is shown in Figure 9. Correlations between any two variables can be formed by joining the variables by the closest time difference. An example, plotting two variables against each other is shown in Figure 10. The output plots and data are available in various formats, including PNG, EPS, HTML, text, CSV, XML, and ROOT.

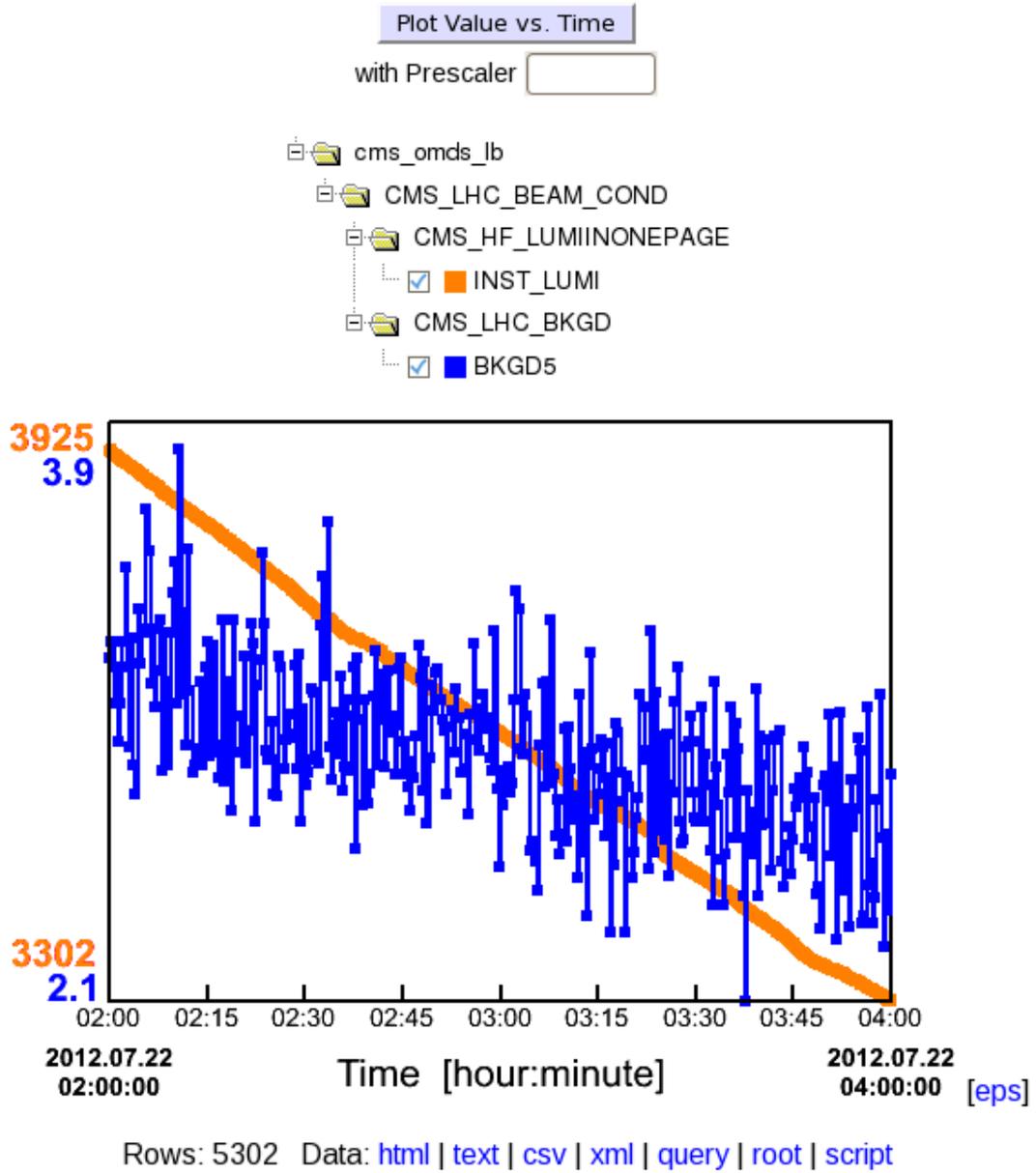

Figure 9. ConditionBrowser – variables versus time for two variables

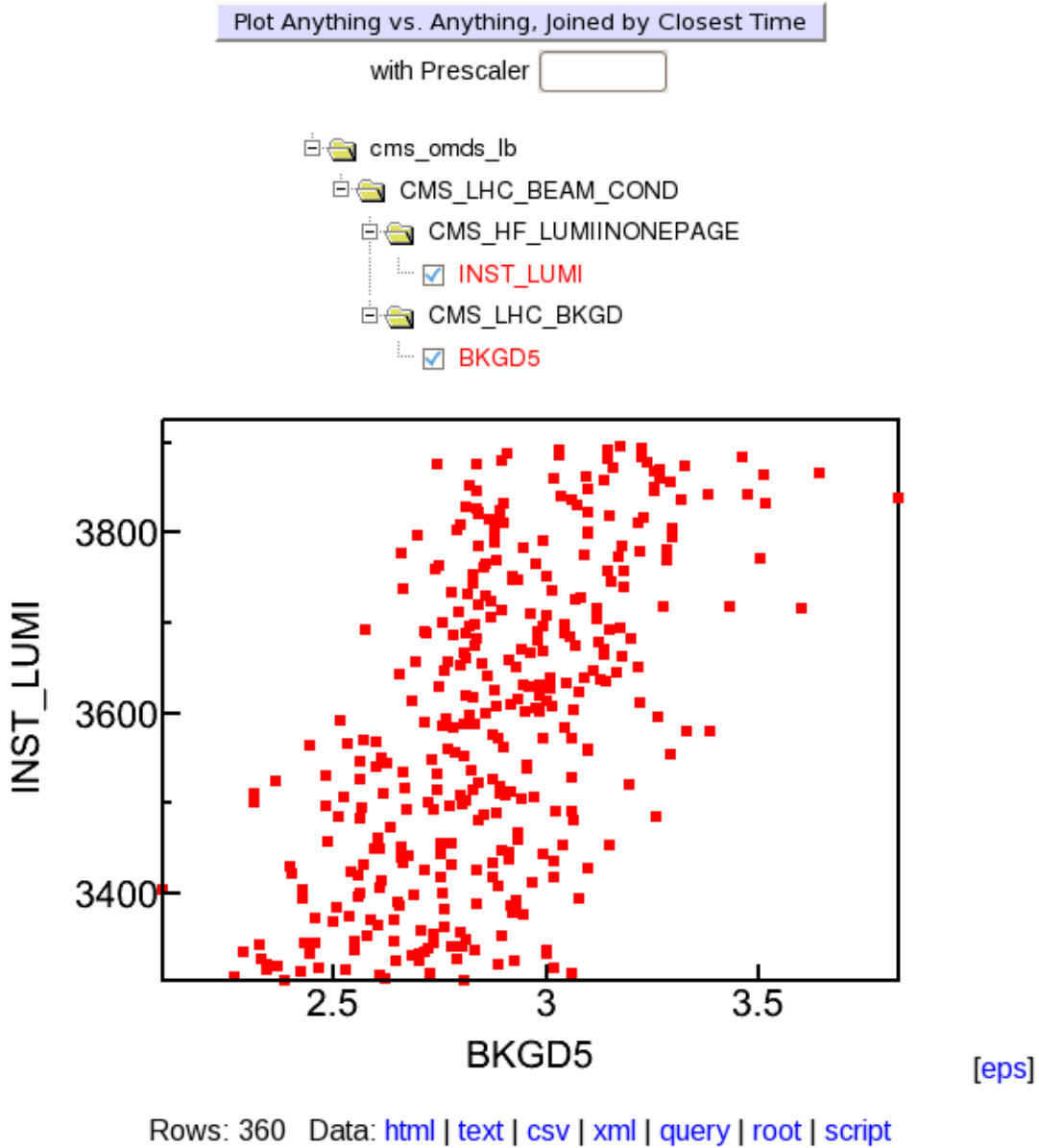

Figure 10: ConditionBrowser – correlation of two variables joined by the closest time difference.

### 3.2 CMS Run Time Logger

The primary goal of the CMS Run Time Logger (RTL) is to help improve CMS data taking efficiency by reporting real time and historical operational efficiency details, for use by the shift crew, operations group, detector experts, and management. In addition to keeping track of down time, live time, and luminosity, it also logs the reasons for down times, allowing users to identify the sources of down time that have the largest impact on data taking. There are four main components to this tool: (1) the underlying Oracle database stores details of downtime events, categories, run periods, detector state, and luminosity. (2) Gaps in data taking, as evidence by the absence of triggers, are automatically detected and logged by the scalers machine described above. (2) A Graphical User Interface (GUI) implemented in Java, is used by the shift crew to enter causes and details of down times. (4) A Web-based reporting tool provides summary plots, pie charts, and sortable tables of down times and efficiencies. An example showing the reasons for down times for the data taking periods for fill 2726 is shown in Figure 11. Figure 12 shows a chart of efficiencies for a set of fills.

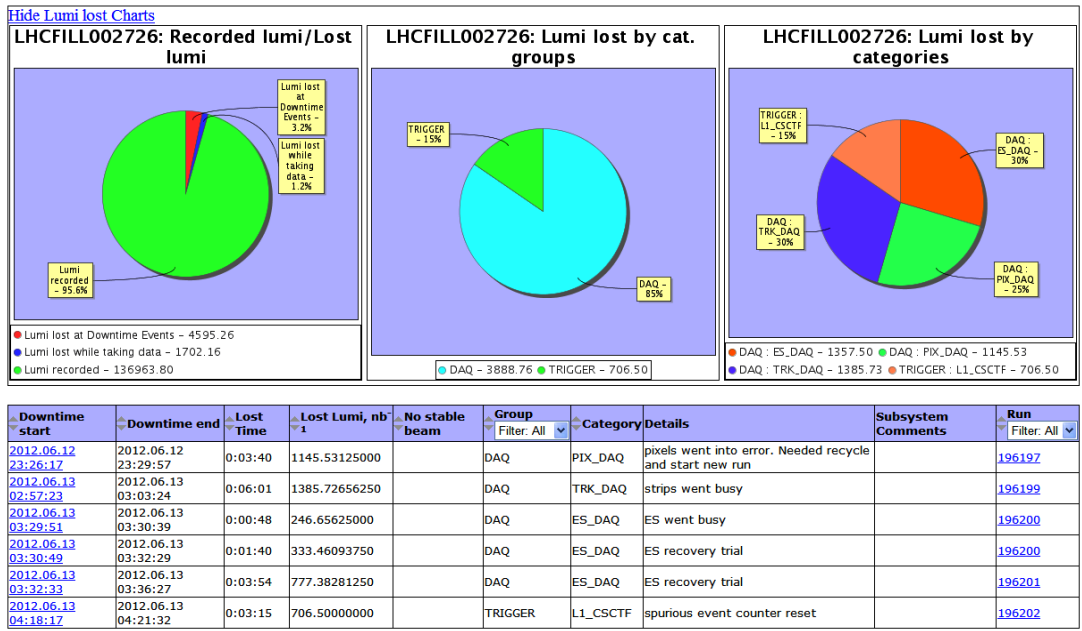

Figure 11. Down time information for a fill from the Run Time Logger

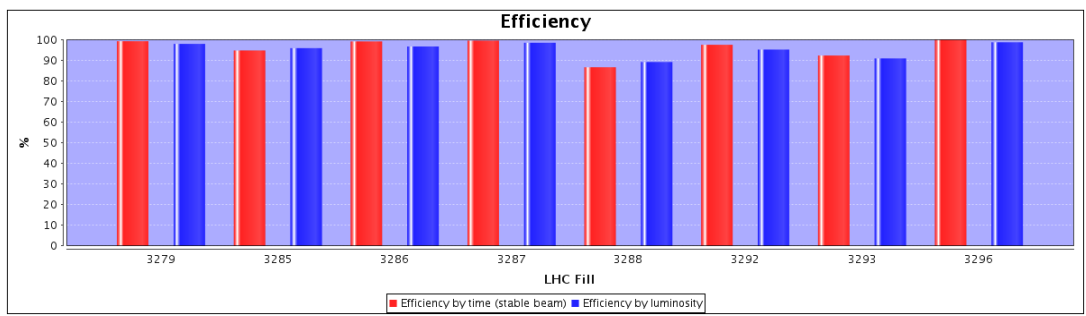

Figure 12. Data taking efficiency vs. fill from the Run Time Logger

### 3.3 CMS Run Registry

The CMS Run Registry (RR) is a tool for tracking the quality of data taking runs for the experiment. This information is used to determine which data taking runs are used by various physics analyses. The Run Registry consists of a suite of applications deployed in the WBM servers. The Online RR provides a user interface and tools for shift personnel to provide the initial quality assessment as runs are taken. The Offline RR provides a user interface and tools for offline shift personnel further evaluating data quality information for runs, this is shown in Figure 13. The User RR application provides a user interface and API for end users to browse, query, and export certification results for use in physics data analysis or other uses.

Figure 13. The CMS Run Registry Tool showing a table of runs, basic information about them, and which detector elements were included. Links are provided to obtain information on what data sets are available for each run, and what detector components were unavailable for which luminosity sections.

### 3.4 WBM Alarm Notification System

The DCS status information logged by the WBM system is used to generate a central alarm in the control room in case of high voltage or other problems in some subsystem. A cronjob script examines this information every minute. Whenever the DCS status of any sub-detector goes off while Stable Beams are declared, a set of WBM DCS warning scripts will raise a sound alarm in the control room. Exceptions are made for the tracker and pixel subsystems at the start and projected end of a fill to allow for turn on/off time. Also, a small pop-up (one per sub-detector) will appear at the shift leader console, informing them about it. The alarm will sound only once, and then be reset after 30 minutes so it will alarm again but only if the problem persists. Monitoring of subsystems with persistent but known problems may be temporarily disabled.

### 3.5 Shift Accounting Tool

At CMS there are more than 70 different types of shift and on-call responsibilities, which are categorized as contributing either to central operations of the experiment or to one of the subsystems. The Shift Accounting Tool (SAT) service enables users to assemble reports of the number and types of shifts taken by people at each collaborating institution during a selected time interval. It is used by individuals, institute group leaders, and run coordinators. Individuals can see details of their own shift contributions. Institution group leaders can see summaries and statistics by shift categories, for individuals in their institution. Run Coordinators and supervisors of shifts can see summaries for particular shift categories, including dates of first shifts and email addresses for shift takers, which are useful for contacting shift takers ahead of their first shifts. Results are reported in tables and plots, and are available in tab-separated text files appropriate for importing into spreadsheets. An important aspect of the tool is that it allows for a comparison between the expected and actual shift contribution by each institution.

**3.6 Snappy eLog Viewer**

In addition to numerical data, substantial information regarding the detector operation is recorded in the form of words. CMS provides an electronic logbook based on Oracle Portal technology [22] for free form entry of this type of information. Entries are categorized by subdetector and operations and are time stamped. Since the data are stored in the online database, many options are possible for displaying and interpreting the logbook content. The Snappy eLog Viewer provides a fast and light-weight method to browse and search the database. Figure 14 shows the primary features of this service including arrangement by 8 hour shift blocks. The user may select various display options and save them in a browser cookie. Although the text entries are not structured data, regular expressions are used to edit the text content and embed links to other WBM services. For example when the word "run or "fill" occurs followed by an integer, a link is created to the appropriate RunSummary or Fill Report pages.

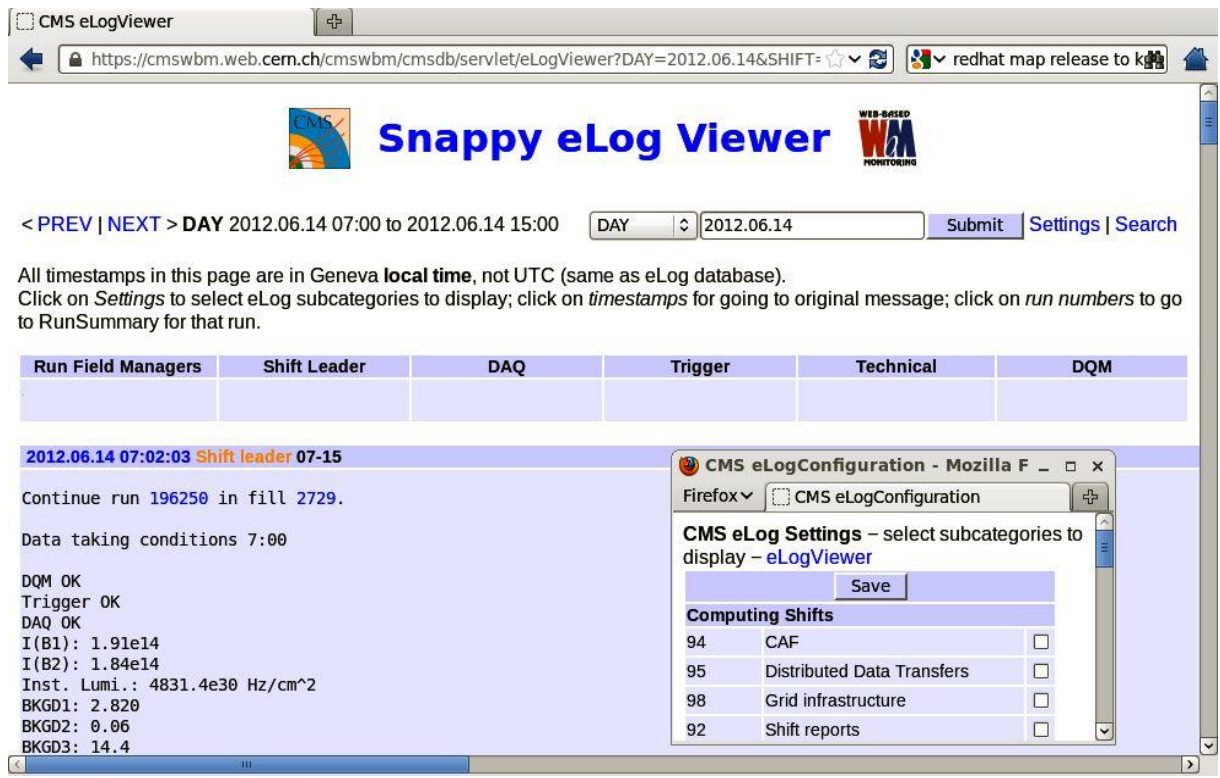

Figure 14. The Snappy eLog Viewer

**4. Subdetector Specific Services**

In addition to the general WBM services described above there are also services specific to individual subsystems within the CMS detector. These deal with voltage and environmental information from the detector control system. Figure 15 shows a display of the RPC system high voltage status. WBM can be used to provide this information correlated with data taking runs. Some subsystems also make detector configuration information available through WBM. These services are created by subsystem personnel, and some have developed a local infrastructure that minimizes the programming required to add new plots or other summary pages. The Tracker subsystem uses template programs based on ROOT to query the database and create plots. The Hadron Calorimeter (HCAL) and Pixel subsystems use "portlets" [23] that provide a modular infrastructure for their displays.

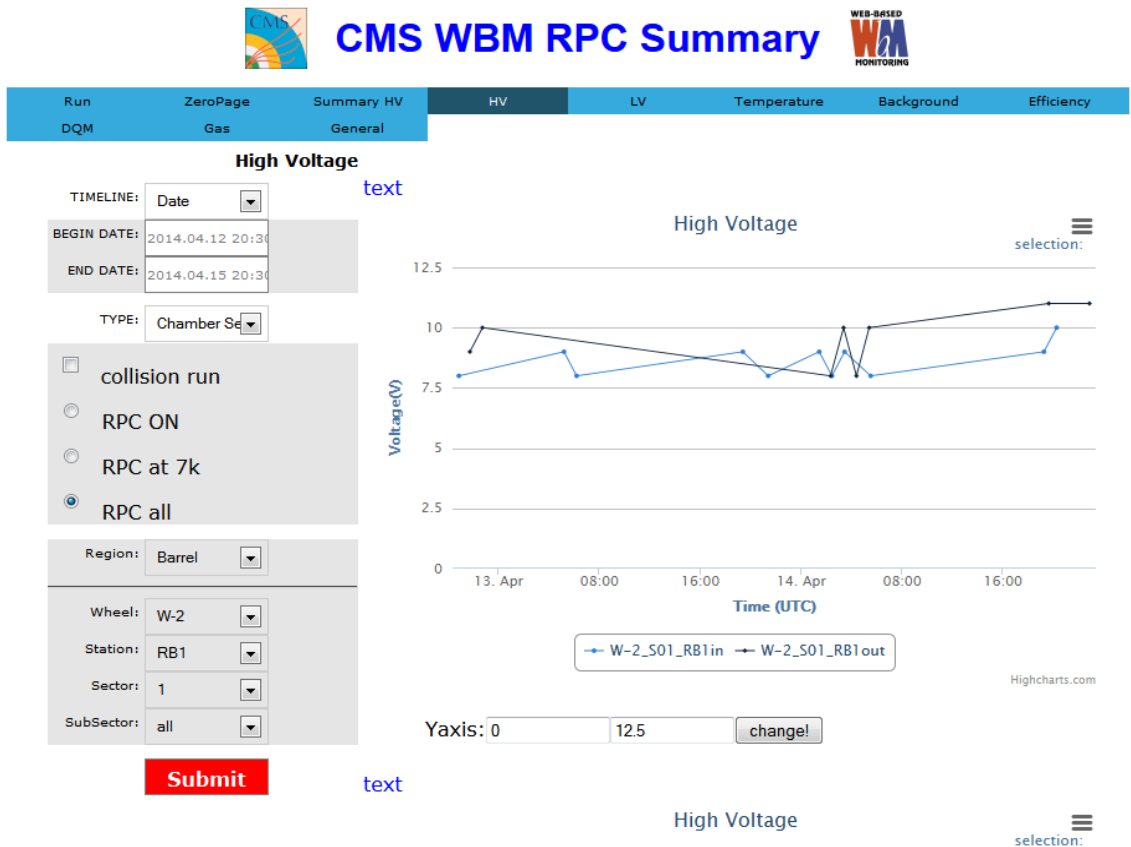

Figure 15. WBM display of the RPC subsystem high voltage status

## 5. Experience

The WBM system proved to be very valuable during the CMS data taking run that ended in early 2013. Information from the WBM system was widely viewed by the collaboration and was an important tool for the run coordination team. During this running period there were an average of approximately 2600 unique visitors and 650,000 page views per month. New features were continually added in response to operational needs. Plots from the system were frequently shown in presentations. The Run Time Logger was very useful in identifying sources of inefficiency in collecting data. Significant effort was then spent in correcting problems causing these inefficiencies. As the luminosity steadily increased during the run, the trigger rate fits were valuable in planning the trigger menu so as to minimize detector dead time. In addition to the core team, a broad range of developers were able to contribute to the subsystem portions. With the large size of the package, significant effort was required to ensure that frequently used data sources were continuously available.

## 6. Future Plans

Currently the LHC is undergoing upgrades and maintenance to allow it to run at higher energy. Operation will resume in early 2015. A number of improvements and adaptations will be made. These include changes in the timing and data acquisition system as well as to specific detector subsystems. More use will be made of HTML5 and the Highcharts plotting package, and the remaining Java applets will be retired. A more general API to access data stored by WBM for either external use or to simplify creating new applications will be generated. The infrastructure will be modified to better implement a REST style architecture. The WBM alarm notification system will be integrated with the central alarm system. It will be possible to include information from the event data quality monitoring (DQM) system in WBM plots. Additional features requested by the run coordination team and the subsystems will be implemented. And general updates of computer hardware, operating systems, and software packages to more recent versions will be done.

## 7. Conclusions

To meet the critical need for monitoring, the WBM project provides a broad suite of tools to convey diverse information on detector operations from many sources. The services are made accessible both locally and remotely, to address the challenges of a large global collaboration. In the first years of collision data taking the Web Based Monitoring was a key element in successfully operating CMS.

## Acknowledgments

We congratulate our colleagues in the CERN accelerator departments for the excellent performance of the LHC and thank all the CMS collaborators, the technical and administrative staffs at CERN and at other CMS institutes for their contributions to the success of the CMS effort. We acknowledge the enduring support for the development and operation of the CMS Web Based Monitoring effort provided by the funding agencies, CERN, the US Department of Energy, and the US National Science Foundation.